\documentclass[a4paper]{article}

\usepackage{latexsym}

\author{S.Albeverio$^{1,2}$, S-M.Fei$^{1,3}$, and P.Kurasov$^{1,4}$}
\title{Gauge fields,
point interactions and few-body problems in one dimension}

\date{}

\begin{document}

\maketitle

$^1$ Institut f\"ur Angewandte Mathematik, Universit\"at Bonn,
D-53115 Bonn

$^2$ SFB 611; BiBoS; CERFIM (Locarno); Acc.Arch., USI (Mendrisio)

$^3$ Dept. of Mathematics, Capital Normal University, 100037 Beijing

$^4$ Dept. of Mathematics, Lund Institute of Technology, 22100 Lund

E-mail: albeverio@uni-bonn.de, fei@uni-bonn.de,
kurasov@maths.lth.se

\begin{abstract}
Point interactions for the second derivative operator
in one dimension are studied. Every operator from this
family is described by the boundary conditions which include
a $ 2 \times 2 $ real matrix with the unit determinant and
a phase. The role of the phase parameter leading to
unitary equivalent operators is discussed in the
present paper. In particular it is shown that the phase parameter
is not redundant (contrary to previous studies) if non stationary
problems are concerned.
It is proven that the phase parameter can be interpreted as
the amplitude of a singular gauge field.
Considering the few-body problem we extend the range
of parameters for which the exact solution can be found
using the Bethe Ansatz.
\end{abstract}


\medskip
{\raggedright {\bf 1} \underline{Introduction.}}

\noindent
The problem of describing point interactions for
one dimensional Schr\"odinger operators attracts considerable
attention by researchers in various disciplines,
since the spectral structure
of such operators can easily be investigated and the
eigenfunctions can be calculated explicitly.
This problem is similar to the one of describing
point interactions for the Laplace operator in $ {\bf R}^3$
analyzed first by F.A.Berezin and L.D.Faddeev \cite{faddeev}.
  From the mathematical perspective the problem in one dimension
is not complicated, since all operators with point interactions are
extensions of one symmetric operator with deficiency indices
$ (2,2) $ and their resolvents can be calculated
using Krein's formula \cite{alb,book,kirsch,gaudin,
jmaa,seba}. The problem of describing all
these operators as singular perturbations of the
free second derivative operator has been solved
in \cite{jmaa}. It has been proven that the four parameter
family of point interactions can be realized as a
combination of a singular potential, a singular density
and a singular gauge field. Therefore all point interactions
for the second derivative operator in one dimension
have been classified.
In particular the one parameter family
of point interactions corresponding to the heuristic Schr\"odinger
operator with a $\delta ' $-potential
has been found,
supporting the conclusions of \cite{griffiths}.
The distribution theory with
discontinuous test functions has been used to
justify the proofs \cite{jmaa,boman}. This approach has been
generalized to include finite rank perturbations of arbitrary
self-adjoint operators \cite{albkur1}.
The relations between this
description of point interactions and the symmetry
properties of the corresponding heuristic Hamiltonians
have been clarified in \cite{adk}.

F.A.B. Coutinho et all \cite{coutinho1}
pointed out that changing one of the parameters describing
these point interactions (the so-called phase parameter)
one gets a family of operators
which are unitary equivalent. In accordance to the classification
given in \cite{jmaa} this parameter corresponds to a
singular gauge field concentrated at the origin. Therefore
it is not surprising that this parameter leads to the
one-dimensional family of unitary equivalent operators,
since the gauge field in one dimension can easily be removed
by a unitary transformation.
In the current paper we further clarify these relations and show
that the phase parameter is not redundant as far as the nonstationary
problem is concerned. These results are generalized for the case of a
system of $ N $ one dimensional particles interacting
via two-body point interactions discussed recently
in \cite{fei,coutinho2,kurasov} following the main ideas of
\cite{mcguire}. It is proven that
the scattering states and bound state functions can be calculated
for a $ 2\left( \! \! \! \!
\begin{array}{c} N \\ 2 \end{array} \! \! \! \! \right) +1 $
parameter family of Hamiltonians, which includes the system of one dimensional particles
with delta two-body interactions as a special case. It is well known that the ground state
eigenfunction for $ N$-particle Hamiltonians with $ \delta$-interactions possesses a boson
like symmetry \cite{lieb1,lieb3,yang}.

\medskip
{\raggedright {\bf 2} \underline{Unitary equivalence of operators with
point interactions}}

\noindent The family of point interactions for the one dimensional
Schr\"odinger operator $ - \frac{d^2}{dx^2} $
can be described by unitary $ 2 \times  2 $ matrices
via von Neumann formulas for self-adjoint extensions
of symmetric operators, since the second derivative
operator restricted to the domain $ C_{0}^\infty ({\bf R}
\setminus \{ 0 \} ) $ has deficiency indices $ (2,2)$
(see \cite{alb,book,chernoff,seba})
(We supposed here for simplicity
that the point interaction is situated
at the origin).
It is more convenient to describe the same family of operators
by certain boundary conditions at the origin imposed
on the functions from the domain of any self-adjoint
extension. In what follows we are going to
consider only the self-adjoint extensions that cannot be
presented as an orthogonal sum of two self-adjoint
operators acting in $ L_{2} (-\infty, 0) $ and
$ L_{2} (0, \infty) $ (the half axes cannot be separated).
Then the boundary conditions describing the operator
have the following form
\begin{equation} \label{bcbc}
\left( \begin{array}{c}
\psi(+0) \\
\psi ' (+0) \end{array} \right)
= e^{i\varphi} \left(
\begin{array}{cc}
a & b \\
c & d \end{array} \right)
\left( \begin{array}{c}
\psi(-0) \\
\psi ' (-0) \end{array} \right),
\end{equation}
where $ \varphi, a,b,c,d \in {\bf R}$,
$ad-bc = 1$. Let us denote by
$ L_{\varphi,a,b,c,d} $
the corresponding self-adjoint operator
defined on the functions from
$ W_{2}^2 ({\bf R} \setminus \{0\})$
satisfying the boundary conditions (\ref{bcbc}).
It has been noted in \cite{coutinho1}
that the operators $ L_{\varphi,a,b,c,d} $
and $ L_{0,a,b,c,d} $ are unitary equivalent.
In fact, consider the function
\begin{equation} \label{eds} \chi (x) =
\left\{
\begin{array}{cc}
1, & x < 0 \\
e^{i\varphi}, & x \geq  0.
\end{array} \right.
\end{equation}
Multiplication by the function $ \chi (x)$ (which has modulus $ 1$)
establishes the unitary equivalence between the operators
$L_{\varphi,a,b,c,d}$ and $L_{0,a,b,c,d}$:
\begin{equation} \label{unittr}
\chi^{*} \, L_{\varphi,a,b,c,d} \, \chi = L_{0,a,b,c,d}.
\end{equation}
The function $ \chi $ is not continuous. Therefore the
domains of the operators $ L_{\varphi,a,b,c,d} $
and $ L_{0,a,b,c,d} $ do not coincide.
The unitary equivalence of the operators implies that all
important physical quantities determined by these
operators are essentially the same as long as the
stationary problem is concerned.

The authors of \cite{coutinho1} concluded that the parameter
$ \varphi $ is redundant, since it produces no interesting
effect.
In what follows we are going to show, however, that this parameter
can play an important role in the study of nonstationary
problems and for these problems interesting effects can be observed.
Consider for example the following equation
\begin{equation}\label{pppp}
L_{\varphi (t),1,0,0,1} \psi (x,t) = \frac{1}{i}
\frac{\partial}{\partial t} \psi (x,t),
\end{equation}
where the phase parameter $\varphi$ depends on the time
(time-dependent gauge field).
The unitary transformation (\ref{unittr})
\begin{equation}
\psi (x,t) \mapsto \phi (x,t) = \chi^{*} \psi (x,t),
\end{equation}
which is time dependent as well,
maps the equation (\ref{pppp}) into the Schr\"odinger equation
with time dependent step potential
\begin{equation}
\left( L_{0,1,0,0,1} - \Theta (x) \varphi' (t) \right)
\phi (x,t) = \frac{1}{i} \frac{\partial}{\partial t} \phi (x,t),
\end{equation}
where $ \Theta = \left\{
\begin{array}{cc}
1, & x \geq 0 \\
0, & x < 0 \end{array} \right. .$
The operator $ L_{0,1,0,0,1} \equiv L $ coincides
with the free second derivative operator on the line.
Consider now the simplest case where the derivative of $ \varphi $
is equal to a negative constant
$ \varphi ' (t) = - h < 0. $
Then the latter equation coincides with the Schr\"odinger equation with
the time independent step potential $ h \Theta (x). $
For this problem one can observe total
reflection of incoming plane waves with small
energies $ 0 < k^2 < h. $
This is in contrast to the free evolution determined
by
$$ L \phi (x,t) = \frac{1}{i} \frac{\partial}{\partial t} \phi (x,t).
$$
This example illustrates the importance of the
phase parameter $ \varphi $ describing the family
of point interactions for the second derivative operator
for nonstationary problems.

\medskip
{\raggedright {\bf 3} \underline{Gauge field and point interactions}}

\noindent
We are going to discuss the physical interpretation of the
phase parameter $\varphi$.
It has been proven in \cite{jmaa} that this parameter
is related to a singular gauge field concentrated
at the origin.
The following heuristic operator has been considered
in \cite{jmaa},
\begin{equation}
L_{\alpha}  =
   \left(  i \frac{\partial}{\partial x} +
\alpha \delta (x) \right)^2 - \alpha^2 \delta (x)^2
   \equiv
- \frac{\partial^2}{\partial x^2}
+ i \alpha \left(
2 \frac{\partial}{\partial x} \delta(x) - \delta^{(1)} (x) \right),
\end{equation}
where by $\delta^{(1)}$ we denote the first (generalized) derivative
of the $\delta$-function.
It has been proven in \cite{jmaa} that the latter
operator coincides with the second derivative operator
$ L_{\alpha} = - \frac{\partial ^2}{\partial  x^2} $
defined on the domain
$$  {\rm Dom} (L_{\alpha} )
   =  \left\{
\psi \in W_{2}^2 ({\bf R} \setminus \{ 0\} :
\left( \begin{array}{c}
\psi(+0) \\
\psi ' (+0) \end{array} \right)
= \frac{2+i\alpha}{2-i\alpha} \left(
\begin{array}{cc}
1 & 0 \\
0 & 1 \end{array} \right)
\left( \begin{array}{c}
\psi(-0) \\
\psi ' (-0)  \end{array} \right) \right\}.
$$
We get boundary conditions from the family
   (\ref{bcbc}) if we put
\begin{equation} \label{edsa}
    e^{i\varphi} =
\frac{2+i\alpha}{2-i\alpha}
\end{equation}
and $ a=d=1, b=c=0. $

The whole four parameter family of boundary conditions for
the second derivative operator has been classified in
\cite{jmaa} in terms of operators with singular
interactions. The singular interactions included a
singular potential, a singular density and a singular
gauge field.
It has been also proven in \cite{jmaa} that a nontrivial
phase in the boundary conditions appears if and only if
the singular gauge field is present.

The fact that the operators with and without singular
magnetic field are unitary equivalent is not
surprising and reminds us about the well known fact
that the one dimensional Schr\"odinger operator
with a smooth gauge field is unitarily
equivalent to the second derivative operator\footnote{Physicists
like to express this fact by saying that ``Gauge field does not
exist in one space dimension"}. In fact let us
consider the one dimensional magnetic Schr\"odinger
equation
\begin{equation}
\left( i \frac{\partial}{\partial x} + a(x) \right)^2
\psi (x,t) = \frac{1}{i} \frac{\partial}{\partial t} \psi (x,t),
\end{equation}
where $ a\in C^1 ({\bf R}) \cap L_{1} ({\bf R}) $.
Then the domain of the operator
$$ L(a) =
\left( i \frac{\partial}{\partial x} + a(x) \right)^2
$$ coincides with the Sobolev space $ W_{2}^2 ({\bf R})$, and the
operator $ L (a) $ is self-adjoint on this domain.
Consider the following new unitary function
$$ \chi_{a} (x) = e^{ i \int_{{-\infty}}^x a(x') dx'} .$$
The operator of multiplication by $ \chi $
establishes the unitary equivalence of the operator $ L(a) $
and the free second derivative operator $ L $
\begin{equation}
\chi^{*}_{a} \, L(a) \, \chi_{a} = L.
\end{equation}
One can observe that for small $ \alpha $ and $ \varphi $,
$ \vert \alpha \vert$, $\vert \varphi \vert \ll 1 $, formula
(\ref{edsa}) reduces to
$$ \varphi=\alpha+O(\alpha^2).$$
This is not surprising, since the operators $ L_{\alpha} $
and $ L (\alpha \delta) $ differ heuristically by a term
proportional to $ \alpha^2 .$
Then substituting $ a(x)=\alpha \delta (x) $ heuristically into
the formula for $ \chi_{a} (x) $ we get
$$ \chi_{\alpha \delta (x)} =e^{i \alpha \Theta (x)}
=\left\{ \begin{array}{cl}1, & x < 0; \\
e^{i\varphi}, & x \geq 0, \end{array} \right. $$
which just coincides with (\ref{eds}).

\medskip
{\raggedright {\bf 4} \underline{Few-body problem}}

\noindent
The unitary transformation discussed in the previous section
for two-body problems with point interactions can easily be
generalized to include the case of few body interactions. Consider a system
of $ N $ equal one dimensional quantum particles with the
coordinates $ x_{i} \in {\bf R}, i=1,2,\ldots,N$
interacting via two body $\delta$-function potentials.
This system is described by the operator
\begin{equation} \label{oper2}
{\cal L}_{N}= - \sum_{i=1}^{N} \frac{\partial^2}{\partial x_{i}^2}
+ \frac{ c}{\sqrt{2}} \sum_{i<j} \delta (x_{i} - x_{j})
\end{equation}
(with $c$ a real parameter)
which is exactly solvable in the sense that all
its eigenfunctions  can be calculated
explicitly as combinations of plane waves
or exponential functions \cite{mcguire,yang}.
The scattering states for this operator can be calculated
using an important analogy with the diffraction problem on
a system of wedges \cite{buslaev,kurasov}.
The book \cite{book} contains the most complete
reference list on this subject.

Let us denote by $ x_{ij} $ the relative Jacobi coordinate associated
with the cluster formed by the particles $i $ and $j$:
$ x_{ij} = \sqrt{2} (x_{i} - x_{j} ) $
and by $ y_{ij} \in {\bf R}^{N-1} $ the relative coordinates
of all other particles with respect to
   the cluster's center of mass.
It is easy to show that the  operator $ {\cal L}_{N} $ coincides
with the Laplace operator
$ {\cal L}_{c} =
- \sum_{i=1}^{N} \frac{\partial^2}{\partial x_{i}^2} $
defined on the domain of functions from
$ \psi \in W_{2}^2 ({\bf R}^{N} \setminus \cup_{i<j} \{ x_{i} = x_{j}
\} ) $
satisfying the following boundary condition on each
hyperplane $ l_{ij} = \{ (x_{ij}, y_{ij} ) \in {\bf R}^{N}: x_{ij} = 0 \} $
\begin{equation} \label{bc2}
\left( \begin{array}{c}
\psi (x_{ij}, y_{ij} ) \\
\displaystyle \frac{\partial \psi}{\partial x_{ij}} \end{array} \right)
\vert_{x_{ij}=+0}
   = \left(
   \begin{array}{cc}
   1 & 0 \\
   c & 1 \end{array} \right)
   \left( \begin{array}{c}
\psi (x_{ij}, y_{ij} ) \\
\displaystyle \frac{\partial \psi}{\partial x_{ij}} \end{array} \right)
\vert_{x_{ij}=-0} .
\end{equation}
Consider now the Laplace operator
$ {\cal L}_{c,a_{ij},\varphi_{ij}} =
   - \sum_{i=1}^{N} \frac{\partial^2}{\partial x_{i}^2} $
defined on the domain of functions from
$ \psi \in W_{2}^2 ({\bf R}^{N} \setminus \cup_{i<j} \{ x_{i} = x_{j}
\} ) $
satisfying the following boundary condition on each
hyperplane $ l_{ij} = \{ (x_{ij}, y_{ij} ) \in {\bf R}^{N}: x_{ij} = 0 \} $
\begin{equation} \label{bc3}
\left( \begin{array}{c}
\psi (x_{ij}, y_{ij} ) \\
\displaystyle \frac{\partial \psi}{\partial x_{ij}} \end{array} \right)
\vert_{x_{ij}=+0}
   = a_{ij} e^{i\varphi_{ij}} \left(
   \begin{array}{cc}
   1 & 0 \\
   c & 1 \end{array} \right)
   \left( \begin{array}{c}
\psi (x_{ij}, y_{ij} ) \\
\displaystyle \frac{\partial \psi}{\partial x_{ij}} \end{array} \right)
\vert_{x_{ij}=-0}
\end{equation}
where $ \varphi_{ij}\in [0,2\pi) $ and $a_{ij}>0$ are real parameters
\footnote{The case of negative $a_{ij}$ can be treated as well by changing
the phase by $\pi$}.
Altogether there are $ 2\left( \! \! \! \! \begin{array}{c}
N \\ 2 \end{array} \! \! \! \! \right) +1  $ real parameters that determine
the operator $ {\cal L}_{c,a_{ij},\varphi_{ij}}$. The operator
$ {\cal L}_{c,a_{ij},\varphi_{ij}} $ is self-adjoint in $ L_{2} ({\bf R}^{N})$
only if all $a_{ij}=1$.
It is easy to show that the operators $ {\cal L}_{c} $
and $ {\cal L}_{c,a_{ij},\varphi_{ij}} $ are similar.
Consider the functions
$$ \chi_{ij} (x_{ij}, y_{ij} ) = \left\{
\begin{array}{ll}
1, &  x_{ij} <0 \\
a_{ij}e^{i\varphi_{ij}}, & x_{ij} \geq 0 .
\end{array} \right.
$$
The following equality establishes the similarity between the operators
${\cal L}_{c,a_{ij},\varphi_{ij}}$ and ${\cal L}_{c}$:
\begin{equation}
\left( \prod_{i<j} \chi_{ij}^{*}\right) \,{\cal L}_{c,a_{ij},\varphi_{ij}}
\, \left( \prod_{i<j} \chi_{ij}\right)
= {\cal L}_{c}.
\end{equation}
The latter equation implies that if the function $ \psi $
is an eigenfunction of the operator $ {\cal L}_{c} $ then
the function
$$ \left( \prod_{i<j} \chi_{ij}\right) \, \psi $$
is an eigenfunction for the operator $ {\cal L}_{c,a_{ij},\varphi_{ij}}.$
It follows in particular that all such operators have the same
negative spectrum. This explains why the ground state calculated
by F.A.B. Coutinho et all \cite{coutinho2} for
the operator ${\cal L}_{c,a,\pm \pi}$ coincides with the result of C.N Yang
\cite{yang}. In the case $a_{ij}\equiv 1$, the operators are unitary
equivalent.

Consider the three body problem. The eigenfunctions for the
operator $ {\cal L}_{c} $ have been first calculated by
J.B.McGuire \cite{mcguire}. There are two types
of eigenfunctions: those generated by incoming plane waves and
those generated by incoming surface waves concentrated near the
lines $ l_{ij}. $ The lines $ l_{ij } $ divide the plane
of zero total momentum into $ 6 $ sectors (see \cite{mcguire}
for details).
The eigenfunctions of the first
type corresponding to the energy
$ E  = k_{1}^2 +k_{2}^2 + k_{3}^3 $
are given in each sector by the set of plane waves
$$ \begin{array}{l}
\psi(x_1,x_2,x_3) =\\
   A_{123} \exp i (k_{1} x_{1} + k_{2} x_{2} + k_{3} x_{3})
+ A_{132} \exp i (k_{1} x_{1} + k_{3} x_{2} + k_{2} x_{3}) \\
+ A_{213} \exp i (k_{2} x_{1} + k_{1} x_{2} + k_{3} x_{3})
+ A_{231} \exp i (k_{2} x_{1} + k_{3} x_{2} + k_{1} x_{3}) \\
+ A_{312} \exp i (k_{3} x_{1} + k_{1} x_{2} + k_{2} x_{3})
+ A_{321} \exp i (k_{3} x_{1} + k_{2} x_{2} + k_{1} x_{3})
\end{array}
$$
with the amplitudes depending on the sector.
These amplitudes can be determined by the boundary
conditions. Similar formula holds for the $N$-body operators,
The function $ \left( \prod_{i<j} \chi_{ij}\right) \, \psi $
is of the same type but satisfies boundary conditions
(\ref{bc3}). It is straightforward to show that the latter
function is an eigenfunction of the operator
$ {\cal L}_{c,a_{ij},\varphi_{ij}}$. Therefore
in addition to the few body system in one dimension
with equal two body delta interactions
there exists a $ 2\left( \! \! \! \!
\begin{array}{c}
N \\
2 \end{array} \! \! \! \! \right) + 1 $-parameter family of few-body problems with point
interactions, that can be solved by McGuire's method. Let us consider the self-adjoint
problem, i.e. when all $a_{ij}=1$. Then the family of $N$-body self-adjoint operators
possessing McGuire solution is described by $ \left( \! \! \! \!
\begin{array}{c}
N \\
2 \end{array} \! \! \! \! \right) + 1 $ real parameters.
 This is in contradiction to
\cite{coutinho2}, where it has been stated that no diffraction
occurs for the scattering states for the three-body Hamiltonian
with point two-body interactions only in the cases
$ e^{i\varphi} = \pm 1. $
The same condition was obtained by investigating the $ N$-body
problem using the Yang-Baxter equation
\cite{fei}. The Yang-Baxter equation is derived for particles having
boson or fermion symmetry. It is obvious that nontrivial phases
$ \varphi \neq 0, \pi $ cannot occur for such particles.

Note that each operator $ {\cal L}_{c,a_{ij},\varphi_{ij}} $
is similar to $ {\cal L}_{c} $.
The operators are unitary equivalent if $ a_{ij} = 1. $
We think that the total many parameter
family of operators can play an important role for the
investigation of $N$-particle Hamiltonians with a time dependent interaction
between the particles.

\end{document}